\documentclass[sigconf]{acmart}
\usepackage{subfigure}
\usepackage[ruled]{algorithm2e}
\usepackage{diagbox}
\usepackage{booktabs}
\usepackage{arydshln}
\usepackage{multirow}
\usepackage{amsmath}
\usepackage{hyperref}

\AtBeginDocument{%
  \providecommand\BibTeX{{%
    \normalfont B\kern-0.5em{\scshape i\kern-0.25em b}\kern-0.8em\TeX}}}

\copyrightyear{2025}
\acmYear{2025}
\setcopyright{acmlicensed}
\acmConference[SIGIR '25]{Proceedings of the 48th International ACM SIGIR Conference on Research and Development in Information Retrieval}{July 13--18, 2025}{Padua, Italy}
\acmBooktitle{Proceedings of the 48th International ACM SIGIR Conference on Research and Development in Information Retrieval (SIGIR '25), July 13--18, 2025, Padua, Italy}
\acmDOI{10.1145/3726302.3729949}
\acmISBN{979-8-4007-1592-1/2025/07}

\begin{document}

\title{Diffusion-based Multi-modal Synergy Interest Network for Click-through Rate Prediction}

\author{Xiaoxi Cui}
\affiliation{
    \institution{Takway.AI}
    \city{Beijing}
    \country{China}
}
\email{cxxneu@163.com}
\authornote{These authors contributed equally to this work.}

\author{Weihai Lu}
\affiliation{
    \institution{Peking University}
    \city{Beijing}
    \country{China}
}
\email{luweihai@pku.edu.cn}
\authornotemark[1]
\authornote{Corresponding author.}

\author{Yu Tong}
\affiliation{%
  \institution{Wuhan University}
  \state{Wuhan}
  \country{China}
}
\email{yutchina02@gmail.com}

\author{Yiheng	Li}
\affiliation{%
  \institution{Shanghai University of International Business}
  \state{Shanghai}
  \country{China}
}
\email{23349096@suibe.edu.cn}

\author{Zhejun Zhao}
\affiliation{%
  \institution{Microsoft}
  \state{Beijing}
  \country{China}
}
\email{anjou1997@gmail.com}

\begin{abstract}
In click-through rate prediction, click-through rate prediction is used to model users' interests. However, most of the existing CTR prediction methods are mainly based on the ID modality. As a result, they are unable to comprehensively model users' multi-modal preferences. Therefore, it is necessary to introduce multi-modal CTR prediction. Although it seems appealing to directly apply the existing multi-modal fusion methods to click-through rate prediction models, these methods (1) fail to effectively disentangle commonalities and specificities across different modalities; (2) fail to consider the synergistic effects between modalities and model the complex interactions between modalities. 

To address the above issues, this paper proposes the Diffusion-based Multi-modal Synergy Interest Network (Diff-MSIN) framework for click-through prediction. This framework introduces three innovative modules: the Multi-modal Feature Enhancement (MFE) Module Synergistic Relationship Capture (SRC) Module, and the Feature Dynamic Adaptive Fusion (FDAF) Module. The MFE Module and SRC Module extract synergistic, common, and special information among different modalities. They effectively enhances the representation of the modalities, improving the overall quality of the fusion. To encourage distinctiveness among different features, we design a Knowledge Decoupling method. Additionally, the FDAF Module focuses on capturing user preferences and reducing fusion noise. To validate the effectiveness of the Diff-MSIN framework, we conducted extensive experiments using the Rec-Tmall and three Amazon datasets.  The results demonstrate that our approach yields a significant improvement of at least 1.67\% compared to the baseline, highlighting its potential for enhancing multi-modal recommendation systems. Our code is available at the following link: https://github.com/Cxx-0/Diff-MSIN.
\end{abstract}




\maketitle

\section{Introduction}
In recent years, the information explosion online has led to information overload, highlighting the importance of personalized recommendation systems and Click-Through Rate (CTR) prediction. Current deep learning CTR models face limitations in capturing evolving user preferences. Recent research addresses this challenge by modeling user behavior sequences, achieving notable progress~\cite{zhou2018deep,zhou2019deep, chen2019behavior, feng2019deep, he2022novel}.  

However, most current approaches only utilize ID-based features in the user's historical behavior sequence, such as item id, category id, etc~\cite{zhou2018deep, pi2020search, chen2021end}. They ignore the valuable textual information (item titles) and visual information (item images) associated with the items. In reality, users are frequently attracted by the titles and images of the products, which in turn impact their click and purchase behaviors. Hence, the fusion of diverse modalities for CTR possesses immense potential\cite{mo2025one}. To be specific, firstly, the complementary relationship between different modalities can provide a more comprehensive expression of user interests~\cite{baltescu2022itemsage}. For example, the text modality can explain the content of the image modality, while the image modality can visually showcase the information from the text modality.
\begin{figure}[htbp]

\centering
\includegraphics[width=0.45\textwidth]{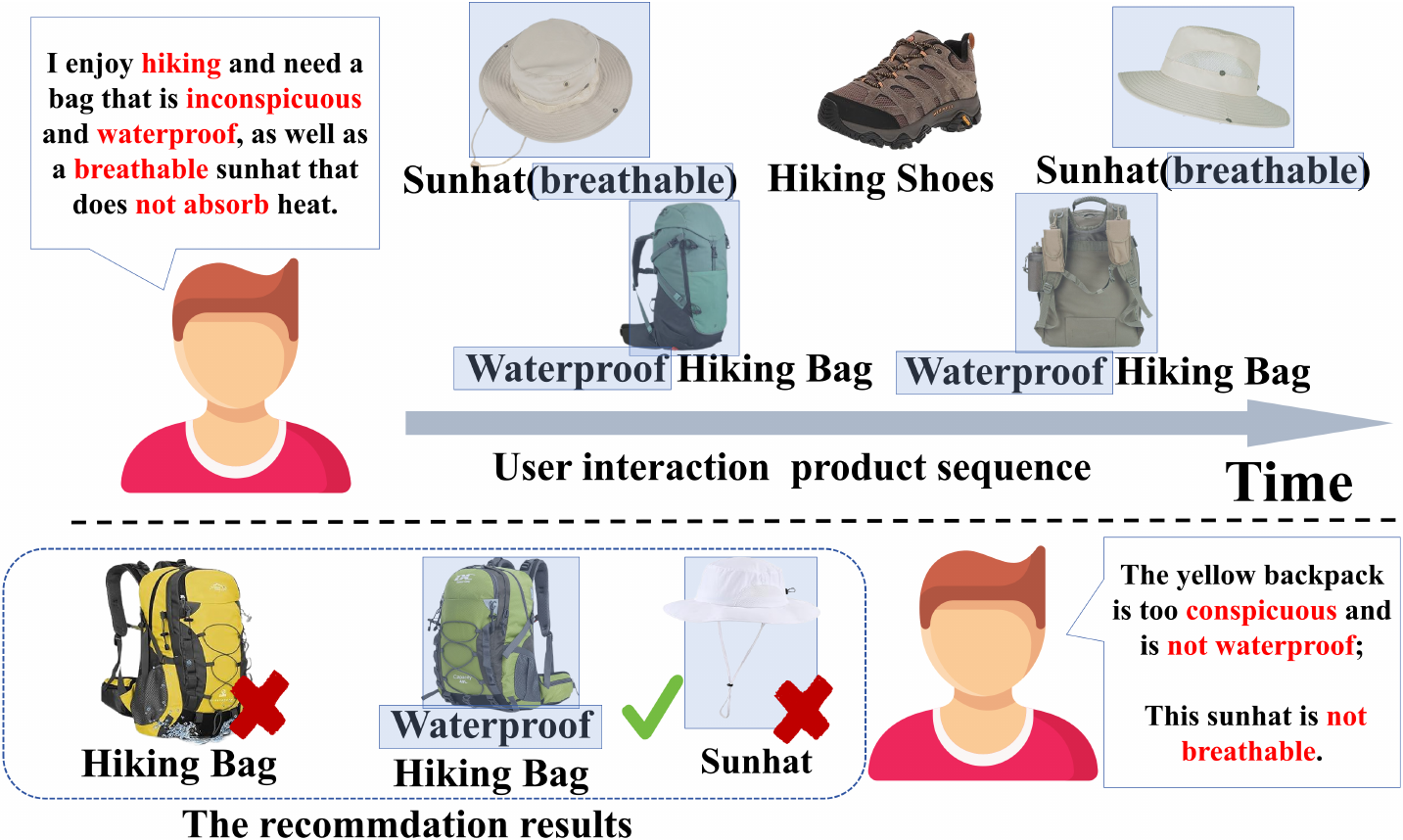}

\caption{User clicks are driven by the synergy of multimodal features (e.g., text and visual). For instance, a hiking bag's "Waterproof" text and "Green" visual features jointly increase click likelihood for users seeking jungle hiking gear; lacking either feature diminishes this likelihood.}
\label{fig:example}
\vspace{-0.5cm}
\end{figure}

The current research on multi-modal recommendation systems
primarily revolves around synergistic filtering and sequence-based recommendation methods. For instance, synergistic filtering methods incorporate multi-modal information into graph neural networks, modeling users and items and leveraging the additional information as edges for data augmentation~\cite{wei2023lightgt}. On the other hand, sequence-based recommendation models utilize multi-modal data as supplementary user features, capturing user interest evolution and behavior patterns~\cite{ji2023online, hu2023adaptive}. These methods have demonstrated superior performance compared to ID modality approaches. 

Nevertheless, the exploitation of multi-modal information in the domain of click-through rate prediction remains largely unexplored. While incorporating applying existing multi-modal fusion techniques directly into existing click-through rate prediction methods (such as DIN~\cite{zhou2018deep} and BST~\cite{chen2019behavior}) may seem appealing. However,

\begin{enumerate}
\item \textbf{Existing methods fail to effectively disentangle commonalities and specificities across different modalities.} The inability to separate common features and modality-specific features in multi-modal data leads to entangled representations. This results in two critical issues: 1) redundant encoding of overlapping information across modalities (e.g., duplicated emphasis on color features in both textual descriptions and product images), and 2) compromised model robustness when handling conflicting signals. As shown in Fig.~\ref{fig:example}, when a user exhibits preference for dark-colored products in most categories while favoring light-colored sunhats specifically, this conflicting pattern may lead the model to erroneously recommend green sunhats. This failure arises from its inability to decouple users' cross-category common preferences  from category-specific requirements. 

\item \textbf{Existing methods fail to consider the synergistic effects between modalities and model the complex interactions between modalities.} Hence, they incorrectly recommend a green breathable sunhat and a non-breathable sunhat to the user. If the recommendation system captures the synergistic relationships between different modalities, it could precisely identify the user’s preferences for light-colored, breathable sunhats and waterproof, green bags, thereby enhancing the precision of its recommendations.
\end{enumerate}
To address the challenges in multi-modal click-through rate prediction, we propose a Diff-MSIN framework that incorporates three innovative modules: MFE, SRC, and FDAF Module. the MFE Module is designed to extract common and special information among different modalities. Inspired by the Progressive Layered Extraction (PLE) framework~\cite{tang2020progressive}, our MFE module utilizes separate expert networks to extract features from text and images, while also employing a shared expert network to capture commonalities across modalities. This approach ensures that different features are not overlooked, resulting in richer and more comprehensive representations of user interests. Inspired by diffusion model~\cite{ho2020denoising}, the SRC module adopts a multi-step synergistic feature interaction approach to capture the synergistic representation. This iterative process empowers the model to progressively refine its comprehension of the relationships among features, capturing both fine-grained and coarse-grained dependencies. By enabling features to interact across multiple time steps, the model can better adapt to the evolving nature of data, resulting in more accurate and robust representations. Meanwhile, Inspired by ~\cite{liu2021noninvasive} the FDAF Module designates ID features as the primary feature and employs an attention mechanism to weight primary feature using auxiliary modality information, effectively reducing noise during the fusion of multi-modal features.

Specifically, our contributions are as follows:
\begin{itemize}
    \item  We propose a general multi-modal user interest modeling framework to model users' cross-modal fused preferences, which can serve as a plugin to enhance performance.
    \item To handle varying representations in multi-modal behavior sequences, the MFE and SRC modules facilitate effective information conversion, capturing synergistic, common, and specific information for efficient multi-modal representation fusion and behavior modeling.
    \item To address noise transmission when fusing synergistic, common, and special features, we designed the FDAF Module. This module improves the quality of fused information by reducing noise and modeling user preferences across modalities, enhancing the modeling process's performance and reliability.
    \item We conduct extensive experiments on real-world datasets to validate the effectiveness of our proposed framework.
\end{itemize}

\section{Related work}
\subsection{Click-through rate prediction}
CTR prediction aims to estimate the probability of a user clicking a candidate item, a task with extensive research. Models like Wide\&Deep~\cite{cheng2016wide} and DeepFM~\cite{guo2017deepfm} are designed to capture low-order feature interactions, whereas DCN~\cite{wang2017deep} and xDeepFM~\cite{lian2018xdeepfm} utilize explicit cross networks for modeling interactions. Approaches like Deep Interest Network (DIN)~\cite{zhou2018deep} and Deep Interest Evolution Network (DIEN)~\cite{zhou2019deep} focus on capturing user interests by modeling behavior sequences, and SIM~\cite{pi2020search} employs a cascaded search paradigm for long-term sequential data. However, these sequence-based models often face limitations in handling very long historical sequences due to computational constraints. Another direction, explored by CIM~\cite{li2021path}, involves modeling users' implicit awareness of candidate and competing items.

\subsection{Diffusion Models for Recommendation}
Recent studies leverage diffusion models for sequential recommendation. DiffuASR~\cite{liu2023diffusion} uses diffusion for data augmentation to combat sparsity and long-tail issues. \cite{wang2024conditional} proposed a conditional denoising diffusion model with a stepwise architecture and novel optimization to mitigate over-smoothing and ranking plateaus. DiffuRec~\cite{li2023diffurec} models items as distributions via diffusion to capture diverse preferences. DiffRec~\cite{wang2023diffusion} learns user interaction generation through denoising, with variants L-DiffRec and T-DiffRec targeting specific challenges. DiffKG~\cite{jiang2024diffkg} integrates diffusion with knowledge graph augmentation and noise filtering. DDRM~\cite{zhao2024denoising} enhances embedding robustness using multi-step denoising. DiFashion~\cite{xu2024diffusion} applies diffusion to personalized outfit recommendation. QARM~\cite{luo2024qarm} offers a quantitative framework for customizing multi-modal information. Separately, SimCEN~\cite{li2024simcen} uses alternate structures and contrastive learning in an MLP to address information loss in CTR models.

\subsection{Multi-modal Recommendation}

Multi-modal approaches are widely studied in recommendation systems to leverage different modalities for capturing user preferences, mainly in collaborative filtering (CF) and sequential recommendation (SR). In CF, research includes using GCNs for modality-specific representations~\cite{wei2019mmgcn}, separating modality-level interests via multi-modal graphs and attention~\cite{tao2020mgat}, improving recommendations with item semantic similarities (LATTICE)\cite{zhang2021mining}, and capturing user preference-item feature correlations\cite{wei2023lightgt}. Sequential recommendation focuses on using multi-modal information to predict the temporal evolution of user interests for personalization~\cite{hou2022towards, hou2023learning, ji2023online, mo2024min}. For Click-Through Rate (CTR) prediction, noting that direct feature fusion is ineffective due to distinct spaces, studies like~\cite{li2020adversarial} and~\cite{xiao2022abstract} employ GANs for feature alignment. Specific models like MAKE~\cite{sheng2024enhancing} address display advertising, and EM3~\cite{deng2024end} targets cold-start/generalization via end-to-end training.

\section{Methodology}

\begin{figure*}[htbp]
\centering
\includegraphics[width=1\textwidth]{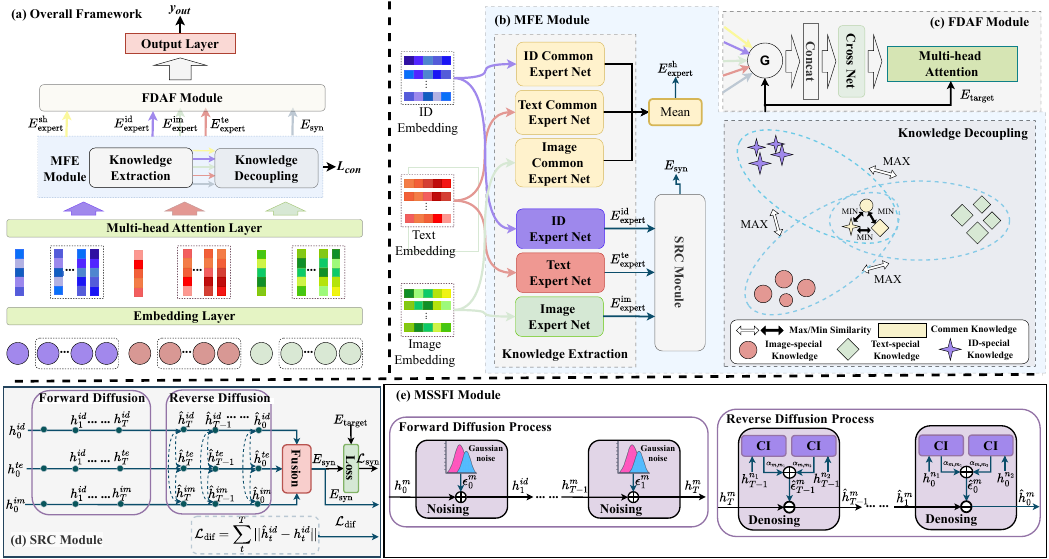}

\caption{(a) The overall framework of our proposed Diff-MSIN framework, which illustrates the forward computation process of different modalities; (b) shows our proposed MFE module, and the SRC module is inclued in MFE module; (c) represents our proposed FDAF module; (d) is the SRC module, and (e) provides detailed information about the MSSFI module.}
\label{fig:framework}

\end{figure*}

This section details our proposed Diffusion-based Multi-modal Synergy Interest Network (Diff-MSIN). We first introduce its inputs and embedding methods, followed by its two main components: the MFE and FDAF Modules. The MFE Module is designed to extract synergistic, common, and special characteristics from different modalities, thereby enhancing feature representations. Concurrently, the FDAF Module employs a non-intrusive fusion approach to reduce noise and adaptively adjusts modal attention weights based on user and target item features. The overall framework and detailed module structures are illustrated in Fig. \ref{fig:framework}.

\subsection{Problem Statement}
\label{sec:pro-sta}

Given a set of users \( U \), the User Profile fields include gender, age, and other relevant attributes. Users' historical behavior sequences are denoted as \( S = \{S^{id}, S^{m1}, S^{m2}, \ldots, S^{mn}\} \), where \( S^{id} \) represents the historical sequence of ID features that includes information such as category, brand, and other identifying attributes (\( S^{id} = \{s^{id_1}, s^{id_2}, \ldots\} \)). The historical sequence \( S^{m1} \) represents the historical sequence of the first modality (\( S^{m1} = \{s^{m1_1}, s^{m1_2}, \ldots\} \)). The target item also possesses multi-modal information, including ID features, text descriptions, and image data. Specifically, the target item can be represented as \( s^t = \{s^{t_{id}}, s^{t_{te}}, s^{t_{im}}\} \). CTR prediction for multi-modal behavior sequences aims to model the relationship between users' historical behaviors, which encompass different modalities, and the target item \( s^t \). 

\subsection{Features Extractor}
\label{sub_sec:Extractor}

In this section, we will discuss the framework inputs and the embedding methods for different modality features. Additionally, we will employ attention mechanisms to initially extract behavioral sequence features from the image and text modalities.

\subsubsection{Framework inputs and features embedding}

The inputs include user features, target item features, and user behavior sequence features. Specifically, the target item features and user's behavior sequence features are further categorized into three types: ID features, text features, and image features.

To generate embeddings for text and images, we utilize TextEncoder and ImageEncoder to process textual and visual information. Among the available options, we opt for the popular CLIP(Contrastive Language-Image Pre-Training) model~\cite{radford2021learning} due to its ability to understand and align cross-modal data, making it suitable for multi-modal modeling. Nevertheless, alternative encoder methods like BERT~\cite{devlin2018bert}, VGG~\cite{simonyan2014very}, and others can also be employed.

\begin{equation}
\label{eq:Embedding}
\begin{aligned}
E_s^{im}, E_{\text{target}}^{im} &= ImageEncoder(S^{im}, s^{t_{im}})\\
E_s^{te}, E_{\text{target}}^{te} &= TextEncoder(S^{te}, s^{t_{te}})
\end{aligned}
\end{equation}

where $S^{im} = \{s_1^{im}, s_2^{im}, \ldots, s_{n_i}^{im}\}$ is the image sequence, where $n_i$ is the length of image sequence. $S^{te} = \{s_1^{te}, s_2^{te}, \ldots, s_{n_t}^{te}\}$ is the text sequence, where $n_t$ is the length of text sequence. $E_s^{im} \in \textbf{R}^{d_i \times n_i}$ and $E_s^{te} \in \textbf{R}^{d_t \times n_t}$ are the image embedding sequence and text embedding sequence, where $d_i$ is the dimension of image embedding and $d_t$ is the dimension of text embedding. $s^{t_{im}}$ is the target image feature and $s^{t_{te}}$ is the target text feature. These features are embedded using a CLIP model, resulting in the target image embedding $E_{\text{target}}^{im} \in \textbf{R}^{d_i}$ and the target text embedding $E_{\text{target}}^{te} \in \textbf{R}^{d_t}$.

\subsubsection{User Interest Modeling}

It is unreasonable to assign equal attention to all items in the user's historical behavioral sequence for both the image and text modalities. For example, when the target item is cloth, items such as pants or clothes in the historical sequence contribute more to the click-through rate prediction. Therefore, we employ attention mechanisms~\cite{zhou2018deep} to capture the attention of each item in the historical sequences of the image and text modalities toward the target item. While other behavioral sequence modeling approaches could be optimized, we chose the most basic one to demonstrate the generalizability of our framework.
\begin{equation}
\label{eq:Attention1}
\begin{aligned}
E_{a,s}^{im} &= Attention_{im}(E_s^{im}, E_{\text{target}}^{im})\\
E_{a,s}^{te} &= Attention_{te}(E_s^{te}, E_{\text{target}}^{te})
\end{aligned}
\end{equation}

Where $Attention_m$ represents the attention mechanism for modality $m$; $E_{a,s}^{im} \in \textbf{R}^{d_i \times n_i}$ and $E_{a,s}^{te} \in \textbf{R}^{d_t \times n_t}$ respectively indicate the weighted embedding sequences of the image and text, which are obtained by applying attention weights.

Due to the potentially long length of the user's historical sequence, it can dramatically enlarge the size of learning parameters. Hence, we sum up the processed representation sequences $E_a^{im} = Sum(E_{a,s}^{im} \in \textbf{R}^{d_i})$ and $E_a^{te} = Sum(E_{a,s}^{te} \in \textbf{R}^{d_t})$) to reduce the parameter size for subsequent processing.

\subsection{Multi-modal Feature Enhancement Module}
\label{sub_sec:Multi-modal Feature Enhancement Module}
\subsubsection{Knowledge Extraction}
According to the description in the introduction, it is crucial to extract both the commonalities and specific characteristics from different modalities. To address this, we propose a multi-modal feature enhancement module. Taking inspiration from PLE~\cite{tang2020progressive}, this module employs two separate Expert Networks to extract text features $E_a^{te}$ and image features $E_a^{im}$. 
\begin{equation}
\label{eq:Embedding_im_te}
\begin{aligned}
E_{\text{expert}}^{im} &= Expert_{im}(E_a^{im})\\
E_{\text{expert}}^{te} &= Expert_{te}(E_a^{te})
\end{aligned}
\end{equation}

Where $Expert_{im}$ and $Expert_{te}$ represent the Expert Networks used to extract image and text features. $E_{\text{expert}}^{im} \in \textbf{R}^{d_e} $ and $E_{\text{expert}}^{te} \in \textbf{R}^{d_e}$ denote the extracted image and text features, where $d_e$ represents the dimension of the output from the Expert Network.

Notably, since existing click-through rate prediction methods like DIN are adept at extracting features from the ID modality, we integrate the embedded click-through rate prediction method as the Expert Network for the ID modality:
\begin{equation}
\label{eq:Embedding_ID}
\begin{aligned}
E_{\text{expert}}^{id} = DIN(S^{id}, s_t^{id})
\end{aligned}
\end{equation}

Afterward, the feature of each modality is fed into the shared expert network, which aims to extract the commonalities across different modalities:
\begin{equation}
\label{eq:Expert_Net}
\begin{aligned}
E_{\text{share}}^{\text{im}} &= Expert^{\text{sh}}_{\text{im}}(E^{\text{im}}_a) \\ 
E_{\text{share}}^{\text{te}} &= Expert^{\text{sh}}_{\text{te}}(E^{\text{te}}_a) \\
E_{\text{share}}^{\text{id}} &= Expert^{\text{sh}}_{\text{id}}(E^{\text{id}}_a) \\
E_{\text{expert}}^{\text{sh}} &= \frac{E_{\text{share}}^{\text{im}}+E_{\text{share}}^{\text{te}}+E_{\text{share}}^{\text{id}}}{3} 
\end{aligned}
\end{equation}

Subsequently, weighted summation of $E_{\text{expert}}^{\text{sh}}$ and the outputs from expert networks are utilized to enhance and complement the feature information from diverse modalities.
\begin{equation}  
\label{eq:Expert_Gate_weight}  
\begin{aligned}
w_m = \sigma_m(E_{\text{expert}}^{m})  
\end{aligned}
\end{equation}
\begin{equation}  
\label{eq:Expert_Gate}  
\begin{aligned}
E^{m} = w_m \odot E_{\text{expert}}^{m} + (1-w_m) \odot E_{\text{expert}}^{\text{sh}}
\end{aligned}
\end{equation}

where $m \in \{im, te, id\}$ is the type of modalities, $\sigma_m$ is the gate network of modality $m$, and $E^m \in \textbf{R}^{d_e}$ represents the features after being weighted by $\sigma_m$.

Overall, the Multi-modal Feature Enhancement Module enables the extraction of both the commonalities and specific characteristics from different modalities, enhancing the capacity to represent multi-modal data.

\subsubsection{Knowledge Decoupling}

To effectively decouple different knowledge domains of user preference, we adopt a contrastive learning strategy. We make the expert tensors $E_{\text{expert}}^{\text{sh}}$, $E_{\text{expert}}^{id}$, $E_{\text{expert}}^{im}$, $E_{\text{expert}}^{te}$ far away from each other in the feature space. This strategy helps the model to better understand and distinguish knowledge in different domains.

Next, we will introduce in detail the contrastive learning method based on cosine similarity to achieve the mutual separation of the five specific tensors and complete knowledge decoupling.

First, the goal of contrastive learning is to learn effective feature representations by adjusting the similarity between sample pairs. In this context, we treat these five tensors as negative sample pairs and expect them to be as far away from each other as possible in the feature space.

For any two expert tensors, the cosine similarity formula is as follows:

\begin{equation}
\cos(E_{\text{expert}}^i,E_{\text{expert}}^j)=\frac{E_{\text{expert}}^i\cdot E_{\text{expert}}^j}{\|E_{\text{expert}}^i\|\|E_{\text{expert}}^j\|},
\end{equation}
where $\|E_{\text{expert}}^i\|$ and $\|E_{\text{expert}}^j\|$ represent the norms of tensors $E_{\text{expert}}^i$ and $E_{\text{expert}}^j$ respectively, which can be obtained by calculating the Euclidean norm of the tensors.

To achieve the mutual separation of the five tensors, we define the loss function as:
\begin{equation}
\mathcal{L}_{con}=\sum_{i}^{M}\sum_{j \neq i}^{M}\cos(E_{\text{expert}}^i,E_{\text{expert}}^j) - \sum_{i}^{M}\sum_{j \neq i}^{M}\cos(E_{\text{share}}^i,E_{\text{share}}^j)
\end{equation}
Where $M = \{id, im, te\}$. By minimizing this loss function using optimization algorithms, we can bring the common features of different modalities closer together while pushing their unique characteristics further apart.

\subsection{Synergistic Relationship Capture Module}

In this section, we introduce our proposed Synergistic Relationship Capture (SRC) Module. The core objective of the collaborative feature extraction module is to synthesize information from different modalities.  Inspired by diffusion models~\cite{ho2020denoising}, we adopt a multi-time-step collaborative approach for multi-modal feature cooperative modeling, aiming to enhance the collaboration between different modalities and the robustness of each individual modality. This is because: (1). the progressive interaction and information exchange between modalities, providing diverse granularities and dimensions for mutual influence and representation updates; (2). Each modality may provide context that helps to denoise another modality. For example, text descriptions can offer details for objects in images, reducing visual noise and improving feature clarity.

\subsubsection{Multi-Step Synergistic Feature Interaction Module}

We design an Multi-Step Synergistic Feature Interaction Module (MSSFI) that incrementally fuses different modality features. Specifically, we perform synergistic feature interaction extraction over $T$ time steps. In each time step $t$, each modality's feature interacts with the features of other modalities to update its representation. The feature vectors for image modality, text modality, and ID modality are represented as $E_{\text{expert}}^{im}$, $E_{\text{expert}}^{te}$, and $E_{\text{expert}}^{id}$, respectively. At the initial state, we define $h^{0}_{im}$ = $E_{\text{expert}}^{im}$,$h^{0}_{te}$ = $E_{\text{expert}}^{te}$,$h^{0}_{id}$ = $E_{\text{expert}}^{id}$.

\textbf{Forward Diffusion Process}
To improve the model's robustness against input noise and modality missingness, we inject random noise into each modality's feature representation after interaction at each time step. The maximum time step is $T$. The noise injection updates at time step $t$ are given by:

\begin{equation}
\hat{h}_{t+1}^{m} = \sqrt{\alpha_t}h_{t+1}^{m} + \sqrt{1-\alpha_t}\epsilon_{t}^{m}
\end{equation}
where $\epsilon_{t}^{m}$ represents the noise vector injected into modality $m$'s feature at time step $t$. $\alpha_t$ controls the degree of noise added at time step $t$. $\hat{h}_{t+1}^{m}$ represents the feature vector of modality $m$ that has been subjected to noise. The noise follows a Gaussian distribution. By introducing noise, the model is encouraged to learn more robust feature representations during training, allowing it to respond better to various disturbances and uncertainties in practical applications.

\textbf{Reverse Diffusion Process}
To facilitate the collaboration among modalities at various granularities and to leverage these cross-modal synergies for noise reduction, we use a cross-modal interaction (CI) function to serve as our denoising function. The update equations for each modality at time step $t$ are as follows:

\begin{equation}
\begin{aligned}
\hat{\epsilon}_{t}^{m} &= \sum_{n \neq m}^{M} \alpha_{m,n} \cdot CI(h_{t}^{m}, h_{t}^{n}) \\
CI(h_{t}^{m}, h_{t}^{n}) &= Attention(h_{t}^{m}, h_{t}^{n}, h_{t}^{n})\\
\hat{h}_{t-1}^{m} &= \frac{1}{\sqrt{\alpha_{t}}} \left( \hat{h}_{t}^{m} - \frac{1 - \alpha_{t}}{\sqrt{1 - \bar{\alpha}_{t}}} \hat{\epsilon}_{t}^{m} \right)
\end{aligned}
\end{equation}

where $M = \{id, im, te\}$, and $CI$ is the cross-modal interaction function. $\alpha_{mn}$ indicates the weight coefficient used to control the extent of information fusion from modality $m$ to modality $n$. These coefficients are learned parameters.

\subsubsection{Synergistic Feature Optimization Based on User Behavior}

To adapt the final synergistic feature representation for downstream tasks, we adjust the loss function according to user click behavior. The synergistic feature $E^\text{syn}_{t}$ dynamically approaches or moves away from the target representation $E_{\text{target}}$ based on clicks. We get $E^\text{syn}$ by fusing $\hat{h}_{0}^{im}$, $\hat{h}_{0}^{te}$, and $\hat{h}_{0}^{id}$ via an MLP:
\begin{equation}
\textbf{E}_\text{syn} = MLP(\hat{h}_{0}^{im}, \hat{h}_{0}^{te}, \hat{h}_{0}^{id}) 
\end{equation}

For positive samples (clicks), we want $E^\text{syn}$ close to $E_{\text{target}}$; for negative samples (non-clicks), we want it far. The loss $\mathcal{{L}}_\text{syn}$ is:
\begin{equation}
\begin{aligned}
    \mathcal{{L}}_\text{syn} &= (1 - y) \cdot \max(0, -1 - \text{cos}(E_{\text{syn}}, E_{\text{target}})) \\&+ y \cdot \max(0, 1 - \text{cos}(E_{\text{syn}}, E^{\text{target}}))
\end{aligned}
\end{equation}
Here, $y$ is click behavior ($y = 1$ for clicks), and $\text{cos}$ is cosine similarity.

\subsection{Feature Dynamic Adaptive Fusion Module}
In this section, we first assign different weights to different modalities based on the target item and user features. Subsequently, we perform denoising fusion on these features.

\subsubsection{Personalized Modality Preferences}
Different users and target items tend to favor specific features or modalities. For example, some users prefer visual images, while certain products emphasize descriptive content. Furthermore, different modalities can complement each other, with images capturing color and style, while text conveys information about fabric and brand for clothing items. Based on these observations, we employ the gate network to weigh different modalities and utilize the cross network for feature interaction.

Specifically, we adaptively calculate the weights for different modalities based on the IDs feature in Eq.~(\ref{eq:Gate_weight}), and subsequently utilize these weights to aggregate the different modalities in the Eq.~(\ref{eq:Gate}). Here, $\sigma$ represents the gate network.
\begin{equation}  
\label{eq:Gate_weight}  
\begin{aligned}
w_{im}, w_{te}, w_{\text{sh}}, w_{\text{syn}} = \sigma(E_{\text{target}}^{id})
\end{aligned}
\end{equation}
\begin{equation}  
\label{eq:Gate}  
\begin{aligned}
&E^{im} = w_{im} \odot E_{\text{expert}}^{im}, E^{te} = w_{te} \odot E_{\text{expert}}^{te} \\ &E^{\text{sh}} = w_{\text{sh}} \odot E_{\text{expert}}^{\text{sh}}, E^{\text{syn}} = w_{\text{syn}} \odot E^{\text{syn}}, 
\end{aligned}
\end{equation}

The resulting processed features are then concatenated as $\hat{E'} = [E^{im}, E^{te}, E^{\text{sh}}, E^{\text{syn}}]$ and inputted into the CrossNet~\cite{yu2021xcrossnet} to facilitate feature interaction between the concatenated auxiliary modality information. In this process, weight $w_c$ and bias $b_c$ parameters are utilized:
\begin{equation}  
\label{eq:Cross_Net}  
\begin{aligned}
E_c = CrossNet(\hat{E'})= \hat{E'}\hat{E'}^T \cdot w_c + b_c + \hat{E'}
\end{aligned}
\end{equation}

$E_c$ is produced by the $CrossNet$ and serves as the auxiliary information.

\subsubsection{multi-modal Feature Fusion}
Directly fusing features from different modalities will introduces noise to the features of each modality during the fusion process. We use attention mechanisms to achieve non-intrusive modal fusion. Specifically, we treat the $E^{id}$ as the primary feature and the $E_c$ as auxiliary features.

The output $E_c$ of the $CrossNet$ and the IDs features $E_{id}$ are inputted into the attention network. The attention network calculates the weights to be applied to $E_{id}$ using the feature information obtained from $E_c$:
\begin{equation}  
\label{eq:Attention2}  
\begin{aligned}
E_{att} = MultiHeadAttention(E_{id}, E_c)
\end{aligned}
\end{equation}

Employing attention mechanisms, we utilize auxiliary features to weight the ID features instead of directly concatenating them. This strategy effectively prevents other modalities from interfering with the ID modality, thereby achieving non-intrusive fusion.

\subsection{Loss}

Since our model is directly embedded into an existing sequential modeling framework, we utilize the loss function provided by the model. For instance, when embedding our framework into DIN, DPN, ETA, and so on, we employ the negative log-likelihood function as the objective function:
\begin{equation}
\begin{aligned}
\mathcal{L}_y = -\frac{1}{N} \sum_{(x,y)\in \mathcal{S}} \left(y \log(y_{out}) + (1-y) \log(1-y_{out})\right)
\end{aligned}
\end{equation}
Here, $\mathcal{S}$ represents the training set with a size of $N$, where each sample $(x,y)$ consists of an input $x$ and a label $y \in {0, 1}$. $y_{out}$ denotes the output of the output layer.
Finally, we combine the $\mathcal{{L}}_{con}$, $\mathcal{{L}}_{syn}$, and $\mathcal{L}_y$ to obtain the final loss $\mathcal{L}$. Here, $w_1$ and $w_2$ is weighting parameters for balancing losses:

\begin{equation}
\label{eq:final_loss}
\begin{aligned}
\mathcal{L} = \mathcal{L}_y + w_1 \cdot \mathcal{{L}}_{con} + w_2 \cdot \mathcal{{L}}_\text{syn}
\end{aligned}
\end{equation}

\section{Experiment}
In this section, we evaluate the Diff-MSIN framework on four public datasets and answer the following research questions:
\begin{itemize}
    \item \textbf{Effectiveness(RQ1).} Can the proposed Diff-MSIN model outperform various state-of-the-art (SOTA) baselines?
    \item \textbf{Generality(RQ2).} Can our proposed framework be applied to different behavioral sequence models and improve their effectiveness?
    \item \textbf{Thoroughness(RQ3).} How do the designs in Diff-MSIN affect the performance of our model?
    \item \textbf{Robustness(RQ4).} How does modifying the parameters in the module affect its effectiveness?
    \item \textbf{Visualization(RQ5).} Does our model effectively capture the synergistic information?
\end{itemize}

\subsection{Experimental Settings}

\subsubsection{Dataset.}We conducted experiments on four real-world datasets: Rec-Tmall, Home, Clothing, and Arts. The Rec-Tmall dataset\footnote{\url{https://tianchi.aliyun.com/dataset/140281}} is sourced from Tmall\footnote{\url{https://www.tmall.com/}}, whereas the Home, Clothing, and Arts datasets are obtained from publicly available sources\footnote{\url{https://jmcauley.ucsd.edu/data/amazon/links.html}}. These datasets are extensively utilized in multi-modal recommendation systems~\cite{xiao2022abstract,li2020adversarial,ji2023online}.For the Rec-Tmall dataset, we utilized product images for visual information representation and product titles for textual information. The ID information included item ID, brand ID, user ID, and seller ID. As for the three Amazon datasets (Home, Clothing, and Arts), we employed product images for visual information, product titles for textual information, and item ID, user ID, and brand ID for ID modalities. For all datasets, we selected user behavior sequences with a minimum length of 5. Additionally, we retained the 50 most recent historical records for each user. For the training and test data, we adopt the same setting as described in~\cite{zhou2018deep, xiao2022abstract}. Table~\ref{tab:datasets} displays the relevant statistical information for each dataset. 

\subsubsection{Evaluation Metrics.}
In our evaluation, we utilize the AUC as a metric to evaluate the quality of prediction results, which is a widely accepted measure in the field of CTR prediction~\cite{fawcett2006introduction}.  Additionally, we introduce the RelaImpr metric, following the methodology described in~\cite{yan2014coupled}, to quantify the relative improvement achieved by different models.

\subsubsection{Implementation Details.} 
The proposed model is implemented using the PyTorch framework\footnote{https://pytorch.org}. To ensure a fair comparison, we utilize our pipeline framework to reproduce all of the baselines, and each baseline model is experimented with multiple times to obtain optimal results. The size of ID modality is set to 16, while image and text modalities embeddings are set to 512 due to the complexity of image and text features compared to IDs. We use a fixed mini-batch size of 1024. When searching for optimal values, we explore learning rates in the set \{10$^{-5}$, 10$^{-4}$, 10$^{-3}$\}, and hidden sizes in the set \{64, 128, 256, 512\}. The weights in Eq.~(\ref{eq:final_loss}) are searched within the range of 0.001 to 0.3. For the contrastive learning component, the dimension of the expert feature representations \( E_{\text{expert}}^{id} \), \( E_{\text{expert}}^{im} \), and \( E_{\text{expert}}^{te} \) is set to 128, consistent with the hidden size explored in our experiments. The diffusion model is configured with a maximum time step \( T \) of 50, and the noise injection parameter \( \alpha_t \) follows a linear schedule from 0.999 to 0.98 over the diffusion steps. The noise vectors \( \epsilon_t^m \) are sampled from a standard Gaussian distribution \( \mathcal{N}(0, I) \). The cross-modal interaction function \( CI \) is implemented using a multi-head attention mechanism with 8 heads and a hidden size of 128. For the expert networks, we use separate MLPs for each modality with a hidden size of 128 and ReLU activation. The gate network \( \sigma_m \) is implemented as a single-layer MLP with a sigmoid activation function to compute the weights \( w_m \) for each modality. To prevent overfitting and optimize performance, we employ an early stopping strategy. Specifically, if the AUC metric does not improve for 10 consecutive epochs, the training process is halted.

\begin{table}[]
\caption{Statistics of Amazon and Rec-Tmall datasets.}

\setlength{\tabcolsep}{4mm}{
\begin{tabular}{c|c|c|c}
\hline
\textbf{Dataset} & \textbf{Users} & \textbf{Items} & \textbf{Interactions} \\ \hline
Home             & 31387        & 64302         & 296428             \\
Clothing         & 64183        & 134064        & 614601             \\
Arts           & 23592        &16340         & 264801               \\
Rec-Tmall         & 72051          & 93466        & 328387             \\ \hline
\end{tabular}}
\label{tab:datasets}

\end{table}

\subsubsection{Baseline Methods.} We compare our framework with state-of-the-art (SOTA) behavioral sequence modeling methods. Traditional and factorization-based CTR models: LR~\cite{mcmahan2013ad}, FM~\cite{rendle2010factorization}; Deep learning-based CTR prediction models: DeepFM~\cite{guo2017deepfm},YoutubeNet~\cite{covington2016deep}, DIN~\cite{zhou2018deep}; Multi-modal CTR Prediction models: LMF~\cite{liu2018efficient}, MTFN~\cite{wang2019matching}, NAML~\cite{wu2019neural}, MARN~\cite{li2020adversarial},GMMF~\cite{xiao2022abstract}, MAKE~\cite{sheng2024enhancing}, EM3~\cite{deng2024end}, QARM~\cite{luo2024qarm}, SimCEN~\cite{li2024simcen}. 

To validate the generality of our framework, we incorporate the following click-through rate prediction models into our framework: DIN~\cite{zhou2018deep}, DPN~\cite{zhang2024deep}, ETA~\cite{chen2021end}, TWIN~\cite{chang2023twin}.

\begin{table*}[]
\caption{AUC on Amazon and Rec-Tmal dataset. Best performances are noted in bold, and the second-best are underlined.}

\setlength{\tabcolsep}{3.5mm}{
\begin{tabular}{c|cc|cc|cc|cc}
\hline
\multirow{2}{*}{\textbf{Method}}  & \multicolumn{2}{c|}{\textbf{Rec-Tmal}} & \multicolumn{2}{c|}{\textbf{Home}} & \multicolumn{2}{c|}{\textbf{Clothing}} & \multicolumn{2}{c}{\textbf{Arts}} \\ \cline{2 - 9} 
                        & \textbf{AUC}          & \textbf{RelaImpr}       & \textbf{AUC}        & \textbf{RelaImpr}     & \textbf{AUC}          & \textbf{RelaImpr}       & \textbf{AUC}         & \textbf{RelaImpr}     \\ \hline
LR(2013)                   & 0.6585       & 0.00\%         & 0.6198     & 0.00\%       & 0.5711       & 0.00\%         & 0.5983       & 0.00\%       \\ 
FM(2010)               & 0.6701       & 1.76\%         & 0.6123     & -1.18\%       & 0.5857       & 2.56\%         & 0.6111      & 2.07\%       \\ \hline
DeepFM(2017)                 & 0.6824       & 3.63\%         & 0.6105      & -1.50\%       & 0.6426       & 12.52\%        & 0.6425      & 7.15\%       \\
YoutubeNet(2016)                  & 0.6873       & 4.37\%         & 0.7325     & 18.18\%      & 0.7403       & 29.63\%        & 0.6852      & 10.82\%       \\
DIN(2018)                 & 0.6839       & 6.89\%         & 0.7383     & 19.12\%      & 0.7061       & 23.64\%        & 0.6664      & 11.02\%       \\ \hline
LMF(2018)                    & 0.7057       & 7.17\%         & 0.7290     & 17.65\%      & 0.6916       & 21.10\%        & 0.6751      & 12.42\%       \\
MTFN(2019)                    & 0.7055       & 7.14\%         & 0.7432     & 20.24\%      & 0.6944       & 21.60\%        & 0.6796      & 13.15\%       \\
NAML(2019)                    & 0.7172       & 8.91\%         & 0.7276     & 17.43\%      & 0.7001       & 22.60\%        & 0.6919      & 15.14\%       \\
MARN(2020)                   & 0.7133       & 8.32\%         & 0.7340     & 18.42\%      & 0.7098       & 24.32\%        & 0.7120      & 18.40\%       \\
GMMF(2022)                   & 0.7124       & 8.19\%         & \underline{0.7428}     & 20.17\%      & 0.7131       & 24.87\%        & 0.7167      & 19.15\%       \\
QARM(2024)  & 0.7152       & 8.49\%         & 0.7338     & 18.40\%      & 0.7101       & 24.34\%        & 0.7135      & 18.67\%       \\
SimCEN(2024)  & 0.7149       & 8.46\%         & 0.7344     & 18.45\%      & 0.7076       & 24.03\%        & 0.7125      & 18.44\%       \\
EM3(2024)                    & \underline{0.7181}       & 9.05\%         & 0.7410     & 19.56\%      & \underline{0.7219}       & 26.41\%        & \underline{0.7199}      & 19.67\%       \\
MAKE(2024)                & 0.7149       & 8.56\%         & 0.7427     & 19.83\%      & 0.7207       & 26.20\%        & 0.7189      & 19.51\%       \\\hline
Diff-MSIN(ours)              & \textbf{0.7270}       & 10.40\%         & \textbf{0.7543}     & 21.70\%      & \textbf{0.7331}       & 28.36\%        & \textbf{0.7312}      & 21.49\%  \\ \hline
\end{tabular}}
\label{tab:AUC_compared_SOTA}

\end{table*}

\begin{table}[]
\caption{AUC on Amazon and Rec-Tmall datasets.}

\setlength{\tabcolsep}{1.3mm}{
\begin{tabular}{c|c|c|c|c}
\hline
\multirow{2}{*}{\textbf{Method}}  & \multicolumn{1}{c|}{\textbf{Rec-Tmal}} & \multicolumn{1}{c|}{\textbf{Home}} & \multicolumn{1}{c|}{\textbf{Clothing}} & \multicolumn{1}{c}{\textbf{Arts}} \\ \cline{2-5} 
                        & \textbf{AUC}           & \textbf{AUC}         & \textbf{AUC}           & \textbf{AUC}           \\ \hline
DIN(2018)                    & 0.6839        & 0.7383      & 0.7061        & 0.6664       \\
DIN+Diff-MSIN           & 0.7270        & 0.7543      & 0.7331        & 0.7312       \\ \hline
ETA(2021)                    & 0.6890        & 0.7389      & 0.7078        & 0.6729       \\
ETA+Diff-MSIN          & 0.7279        & 0.7550      & 0.7334        & 0.7321       \\ \hline
TWIN(2023)                 & 0.7095        & 0.7391      & 0.7106        & 0.6763       \\
TWIN+Diff-MSIN           & 0.7293        & 0.7576      & 0.7362        & 0.7355       \\ \hline
DPN(2024)             & 0.7142        & 0.7407      & 0.7109        & 0.6879       \\
DPN+Diff-MSIN          & 0.7301        & 0.7583      & 0.7392        & 0.7369       \\ \hline
\end{tabular}}
\label{tab:generality experiment}

\end{table}

\subsection{Performance Comparison (RQ1)}

To validate the effectiveness of our proposed model, we conducted experiments on four different datasets, embedding the classical behavior sequence model DIN into our framework. The results, shown in Table~\ref{tab:AUC_compared_SOTA}, compare Diff-MSIN with baseline models. (1) \textbf{Our Method Outperforms All Baselines}. We compared different methods on the Rec-Tmal, Home, Clothing, and Arts datasets, evaluating performance using the AUC metric. The results indicate that Diff-MSIN achieved the highest AUC values across all datasets, demonstrating its effectiveness. (2) \textbf{Our Method Models the Synergy Between Modalities Better Than Others}. EM3, MAKE, and GMMF consider the differential features of different modalities, but they fail to model the synergy relationship, which may misrepresent user preferences based on specific modal features. In contrast, Diff-MSIN effectively captures both common characteristics and synergy information across modalities. (3) \textbf{Multi-modal Methods Outperform Non-multi-modal Methods}. Overall, multi-modal methods  superior performance compared to non-multi-modal methods. By leveraging multiple modalities, they capture more comprehensive information, resulting in more accurate predictions. Diff-MSIN, in particular, excels due to its thorough modeling of various modal features and their interactions.

\subsection{Generalizability Study (RQ2)}
To investigate the applicability of Diff-MSIN to different behavioral sequence models, we incorporated various models into our study. The obtained results are presented in Table~\ref{tab:generality experiment}, and based on these experimental findings, we draw the following conclusions:

\begin{itemize}
    \item Diff-MSIN demonstrates a high degree of generalizability, as it consistently yields improved performance across different behavioral sequence models.  This suggests that the proposed approach can effectively enhance the effectiveness of various models in capturing and modeling behavioral patterns.
    \item The generalizability of Diff-MSIN is evident across different domains and datasets. We observed consistent improvements in performance across diverse datasets, reinforcing the versatility and applicability of our approach.
\end{itemize}

In summary, our study demonstrates that Diff-MSIN exhibits strong generalizability to different behavioral sequence models. The consistent performance improvements and statistical significance validate the effectiveness of our approach in enhancing the effectiveness of various models across different domains and datasets.

\begin{table*}[]
\caption{AUC on Amazon and Rec-Tmall datasets in ablation experiments.}

\setlength{\tabcolsep}{3.5mm}{
\begin{tabular}{c|cc|cc|cc|cc}
\hline
\multirow{2}{*}{\textbf{Method}} & \multicolumn{2}{c|}{\textbf{Rec-Tmal}} & \multicolumn{2}{c|}{\textbf{Home}} & \multicolumn{2}{c|}{\textbf{Clothing}} & \multicolumn{2}{c}{\textbf{Arts}} \\ \cline{2-9} 
                                 & \textbf{AUC}    & \textbf{RelaImpr}    & \textbf{AUC}  & \textbf{RelaImpr}  & \textbf{AUC}    & \textbf{RelaImpr}    & \textbf{AUC}   & \textbf{RelaImpr}  \\ \hline
w/o MFE and SRC                          & 0.7172            & -0.82\%              & 0.7389        & -1.94\%            & 0.7237          & -1.11\%              & 0.7354         & -1.96\%            \\ \hline
w/o FDAF                         & 0.7190           & -0.57\%              & 0.7474        & -0.81\%            & 0.7295          & -0.31\%              & 0.7403         & -1.31\%            \\ \hline
w/o SRC                         & 0.7169           & -0.85\%              & 0.7433        & -1.35\%            & 0.7278          & -0.55\%              & 0.7399         & -1.36\%            \\ \hline
w/o MFE,SRC,FDAF                 & 0.7138          & -1.29\%              & 0.7328        & -2.75\%            & 0.7182          & -1.86\%              & 0.7325         & -2.35\%            \\ \hline
\end{tabular}}
\label{tab:ablation experiment}

\end{table*}

\subsection{Ablation Study (RQ3)}
In this section, we conduct experiments to assess the impact of various modules on the performance of recommender systems:

\textbf{w/o MFE, SRC and FDAF}: We remove the MFE, SRC, and FDAF modules.

\textbf{w/o FDAF}: We exclude the FDAF module.

\textbf{w/o MFE}: We eliminate the MFE module.

\textbf{w/o SRC} we remove the SRC modules.

The experimental results in Table~\ref{tab:ablation experiment} demonstrate the impact of different modules on the performance of the recommendation system. (1) Without the MFE, SRC and FDAF modules, the system achieves the lowest AUC values on all datasets, indicating that these modules play a crucial role in improving recommendation accuracy. (2) Removing the FDAF module while keeping the MFE module leads to slight improvements in AUC values for all datasets. This suggests that the MFE and SRC module can capture and enhance the synergistic, common, and special characteristics among different modalities, contributing to better recommendation performance. (3) Similarly, removing the MFE and SRC while retaining the FDAF module also yields improvements in AUC values. This indicates that the FDAF module effectively captures user preferences and reduces fusion noise. (4) Overall, the experimental results highlight the significance of taking into account modal commonality, specificity, and synergic relationships when modeling multi-modal data for recommendation systems. Furthermore, accurately capturing user preferences across different modalities and minimizing noise during the fusion process can further improve the accuracy of recommendations.

\subsection{In-depth Analysis(RQ4)}

\begin{figure}[htbp]
\centering
\includegraphics[width=0.35\textwidth]{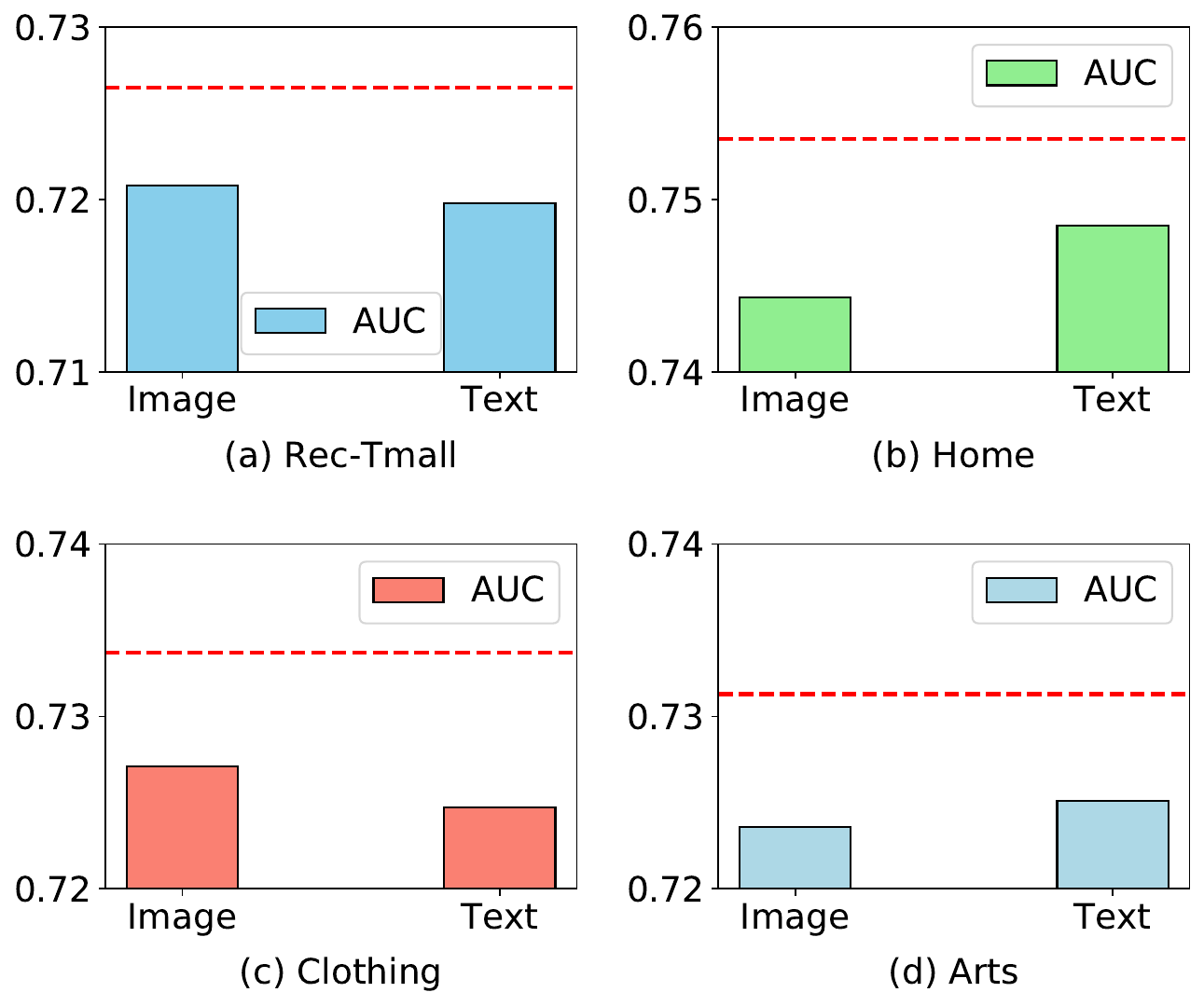}
\vspace{-0.3cm}
\caption{AUC for removing different modalities}
\vspace{-0.5cm}
\label{fig:remove_modality}
\end{figure}

\textit{Effect of removing different modalities.} To validate the contribution of different modalities to click-through rate prediction, we performed experiments by removing each modality individually. Fig.\ref{fig:remove_modality} shows that removing either the text modality or the image modality leads to a decrease in AUC, indicating that both modalities contribute to the accuracy of click-through rate prediction. The influence of removing different modalities varies across different datasets, suggesting that different types of products may have preferences for different modalities.

\begin{figure}[htbp]
\centering
\includegraphics[width=0.40\textwidth]{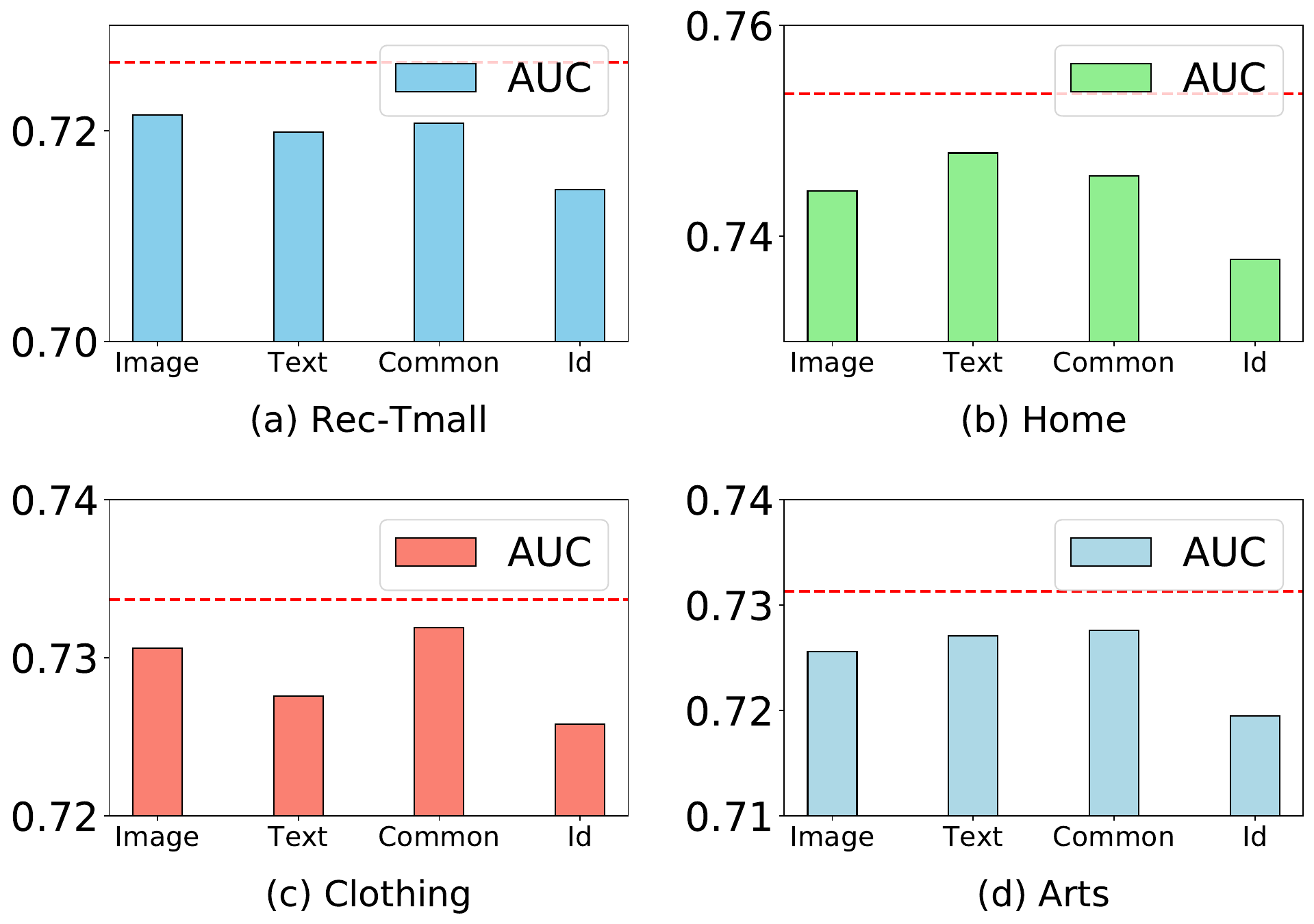}
\caption{AUC for removing different expert net}
\vspace{-0.5cm}
\label{fig:remove_expert_net}
\end{figure}

\textit{Effect of removing different characteristics.} As depicted in Fig.~\ref{fig:remove_expert_net}, the removal of different characteristics consistently leads to a decline in model performance, and this decline is positively correlated with the importance of the modalities. As observed in Fig. ~\ref{fig:remove_modality}, the removal of characteristics corresponding to more critical modalities results in a more significant performance drop. This is because expert net are adept at extracting modal-specific features, thereby enhancing modal representations. Additionally, the removal of common and synergistic characteristics also causes a decrease in model performance. This can be attributed to the characteristics' ability to harness the consistency and synergy between multiple modalities, extracting more universally shared representations and synergistic characterizations, while filtering out redundant information across modalities. 

\textit{Effect of the setting of time step $T$.} A small $T$ (e.g., 5) limits iterations, hindering cross-modal integration and capturing only superficial features (AUC 0.7203). Conversely, a large $T$ (e.g., 20) causes overfitting, reducing interaction diversity and generalization (AUC 0.7192). `T = 12` strikes a balance, enabling sufficient modal interaction without overfitting, achieving the best AUC (0.7264). Thus, $T$ significantly impacts multi-modal collaboration and accuracy, with an optimal range around 10-15.

\textit{Effect of multi-modal embedding sources.} 
To validate the influence of different embedding methods on our model, we utilized a Transformer to embed both text and image modalities.  We also employed BERT for text embedding and VGG for image embedding.   Table~\ref{tab:embedding_method} shows that the variations in performance among the different embedding methods were minimal, with CLIP outperforming the others.  This could be attributed to the inherent strengths of CLIP, including its capacity to capture comprehensive semantic information and effectively align text and image representations.

\begin{figure}[htbp]
    \vspace{-0.1cm}
    \centering
    \begin{minipage}[b]{0.23\textwidth}
        \centering
        \includegraphics[width=1\textwidth]{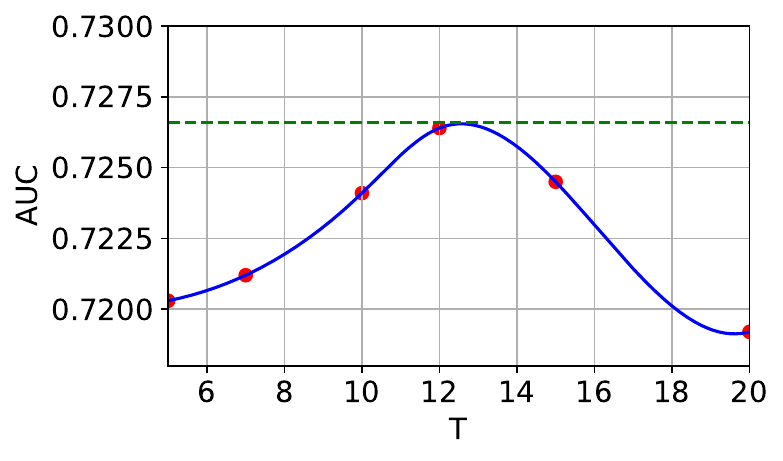}
        
        \caption{The influence of $T$ setting on AUC}
        \label{fig:effect_t}
    \end{minipage}
    \hfill
    \begin{minipage}[b]{0.23\textwidth}
        \centering
        \includegraphics[width=1\textwidth]{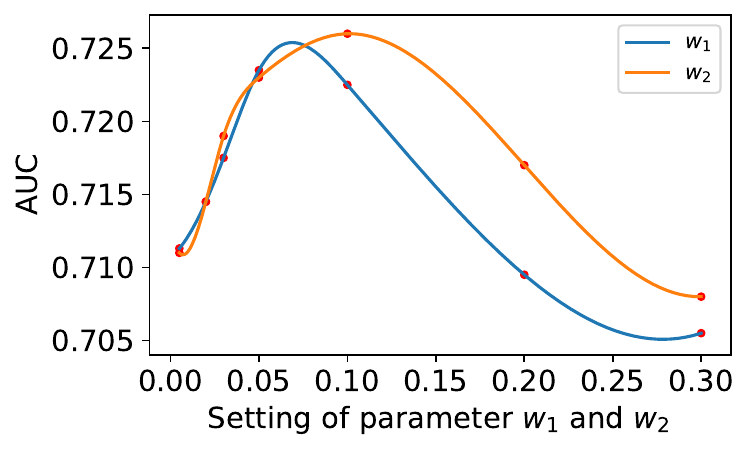}
        \vspace{-0.2cm}
        \caption{The influence of $w_1$ and $w_2$ setting on AUC}
        \label{fig:effect_w}
    \end{minipage}
    
\end{figure}

\textit{Effect of $w$ in Eq.~(\ref{eq:final_loss})}
To validate the impact of the weight parameter $w$ in the loss function Eq.~(\ref{eq:final_loss}), we conducted experiments by varying $w_1$ and $w_2$ within the range of $[0.0001, 0.05]$ on Amazon Home. Fig.\ref{fig:effect_w} illustrates the results, indicating that a smaller $w_1$ value leads to a reduced ability of the model to differentiate between different modality-specific characteristics, reducing the effectiveness of the MFE module and resulting in a decrease in AUC. A smaller $w_2$ causes the model to reduce its ability to extract the synergistic relationships of different features, thus decreasing the AUC. Conversely, a larger $w_1$ or $w_2$ value, which emphasizes distinguishing modality-specific characteristics and synergistic characteristics over improving recommendation accuracy, leads to a substantial decline in AUC.

\begin{table}[]
\caption{AUC for Different Embedding Methods}

\setlength{\tabcolsep}{3mm}{
\begin{tabular}{c|ccc}
\hline
\multirow{2}{*}{\textbf{Dataset}} & \multicolumn{3}{c}{\textbf{AUC}}                                                                   \\ \cline{2-4} 
                                  & \multicolumn{1}{c|}{\textbf{CLIP}} & \multicolumn{1}{c|}{\textbf{Transformer}} & \textbf{BERT+VGG} \\ \hline
Home                              & \multicolumn{1}{c|}{0.7537}        & \multicolumn{1}{c|}{0.7510}               & 0.7496            \\
Clothing                          & \multicolumn{1}{c|}{0.7321}        & \multicolumn{1}{c|}{0.7315}               & 0.7292            \\
Arts                            & \multicolumn{1}{c|}{0.7507}        & \multicolumn{1}{c|}{0.7512}               & 0.7500            \\
Rec-Tmall                          & \multicolumn{1}{c|}{0.7267}        & \multicolumn{1}{c|}{0.7221}               & 0.7232            \\ \hline
\end{tabular}}
\label{tab:embedding_method}
\end{table}

\begin{table}[]
\caption{Time Efficiency Comparison on Rec-Tmall Dataset}

\setlength{\tabcolsep}{2mm}{
\begin{tabular}{c|c|c}
\hline
\multirow{2}{*}{\textbf{Method}} & \textbf{Training Time} & \textbf{Inference Time} \\ 
                & \textbf{per Epoch (s)} & \textbf{per Prediction (s)} \\ \hline
SimCEN & 77.0 & 0.18 \\
EM3 & 78.7 & 0.17 \\
MAKE & 89.5 & 0.33 \\
\textbf{Diff-MSIN (Ours)} & 97.3 & 0.36 \\ \hline
\end{tabular}}
\label{tab:time_efficiency}

\end{table}

\subsection{Time Efficiency Experiment} We compared Diff-MSIN's training and inference times with baselines on the Rec-Tmall dataset under identical hardware (Table~\ref{tab:time_efficiency}). Diff-MSIN's longer training time, due to extra complexity from denoising and multi-expert layer, is justified by performance gains. Diff-MSIN is comparable to complex MAKE, showing it doesn't significantly increase inference time among multi-modal CTR models.

\subsection{Case Study(RQ5)}
\begin{figure}[htbp]
\centering
\includegraphics[width=0.5\textwidth]{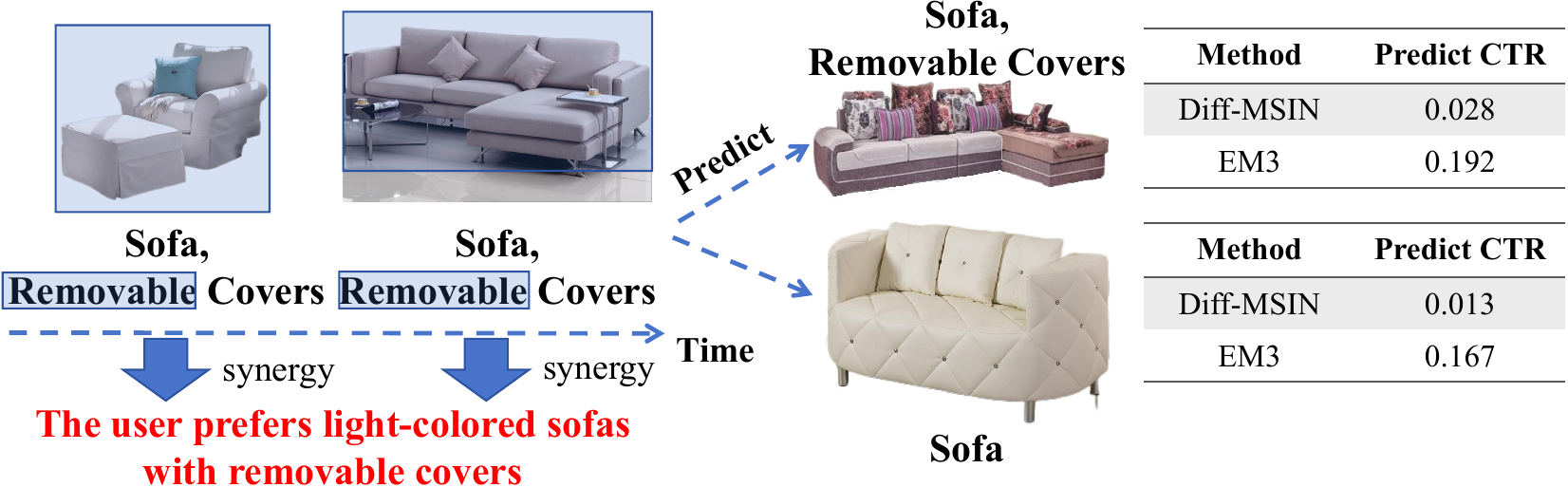}

\caption{Case study}

\label{fig:cos_impr}
\end{figure}
In Fig.~\ref{fig:cos_impr}, on the left is the user history click sequence. The task is to predict the click-through rate (CTR) of the two target items based on the historical sequence. The result shows that the CTR predicted by Diff-MSIN is lower than that of EM3. Since the two target items were not clicked at the next moment, the Diff-MSIN prediction was more accurate. This is because Diff-MSIN effectively captures the synergistic preference in the user's historical sequence for light-colored sofas with removable covers. However, EM3 considers the features of text or pictures separately and wrongly believes that the user might click the items, reducing the prediction accuracy.

\section{Conclusion}
This paper addresses the challenges in effectively incorporating and fusing information from diverse modalities. We propose the Diff-MSIN framework. The Diff-MSIN contributes to enhancing the representation of modalities by capturing synergistic, common, and special information from different modalities, and reducing noise during fusion. Experimental results on four datasets validate the effectiveness of Diff-MSIN, demonstrating a significant improvement over the baseline approach. 

\newpage
\bibliographystyle{ACM-Reference-Format}
\balance
\bibliography{sample-base}

\appendix

\end{document}